# Transients with time-independent currents

Er'el Granot and Avi Marchewka

*Department of Electrical and Electronics Engineering, Ariel University Center of Samaria, Ariel, Israel*

**Abstract**

It is shown that when the initial particles probability density is discontinuous the emerging currents appear instantaneously, and although the density beyond the discontinuity is initially negligible the currents there have a finite value. It is shown that this non-equilibrium effect can be measured in real experiments (such as cooled Rubidium atoms), where the discontinuity is replaced with finite width (hundreds of nanometers) gradient.

**Introduction**

In principle there should not have to be a direct link between the carriers' current and the carriers' density. For example, in metals since the carriers (electrons) surroundings, i.e., the ions, have a positive charge, which cancels out the carriers' negative charge, the overall charge *density* is practically zero, while the current can be arbitrarily large.
However, whenever there is only one type of carriers it seems contradictory to obtain finite current with zero density.

The reason that we usually expect from the current to be proportional to the density is because we are accustomed to stationary-state current. Stationary- state and close-to-equilibrium conductance are easy to handle and in many scenarios these are excellent approximation, which led to many important insights about the quantum world. One of these finding is the Landauer equation[1], whose success made it the golden standard of conductance.

However, whenever there is current – something has to change, and if something has to change then the state is no longer stationary. Right after the current is turned on the stationary-state approximation is evidently useless.

Mosinsky, motivated by the time diffraction of Fraunhofer diffraction, constructed a quantum shutter[2]. His shutter model consists the free propagation of a step function as an initial condition. Many works have follow and been inspired by Moshinsky's work[3], who showed diffraction in time . Independently of the quantum shutter and motivated by fundamental quantum problems (e.g., absorption boundary, localization of wavefunctions and the quantum measurements theory) we investigated the short time propagation of generic initially singular wavefunctions [4-5]. The generic formula was applied in numerous scenarios, some of which are quite counterintuitive. Although, from a formal point of view, quantum mechanics does support singular wavefunction, and thus all the result of singular case are valid [6], from the practical point of view singular wavefunctions may and does raise suspicion. In order to attack this problem it has been shown that if the initially singular wavefunction is replaced with a high gradient continuous one, most of the interesting effects are still valid under certain restrictions [5]. Therefore, singularity is a matter of measurements and the singular wavefunctions can be the trigger for many fruitful investigations. We use the same rational in this work. We first calculate the current of an initially singular wavefunction, and then, motivated by the peculiar results, calculate it for a continuous with high gradient initial condition. We show that in both cases a constant current appears for a certain interval of time. A possible realization of the effect is then given.

The Moshinsky dynamics can be generalized to arbitrary initially singular wavefunction [4], i.e., if initially the wavefunction is

$$\psi(x, t=0) = \varphi(x)\Theta(-x) \tag{1}$$

where $\varphi(x)$ is any analytical function and $\Theta(x)$ is the Heaviside step function then

$$\psi(x>0, t>0) = \frac{1}{2}\exp\left(i\frac{x^2}{2t}\frac{m}{\hbar}\right) w\left(\frac{-(\hbar/m)it\partial/\partial y - x}{\sqrt{-2it\hbar/m}}\right)\varphi(y,0)\bigg|_{y=0} \qquad (2)$$

where $w$ is the Faddayeva function [7].
.

It was shown that at short times universalities emerges so the specific shape of the initial wavefunction is of no importance at short time provided the singularity is similar.

**The singular case**: To simplify the matter, we will focus on a simple step function. As was stressed above, the differences to other wavefunctions are miniscule at short times. Thus, we choose $\varphi(x) = \sqrt{n}$, where $n$ is a constant probability density.

In this case, the solution resembles Moshinsky's shutter:

$$\psi(x>0, t>0) = \frac{1}{2}\exp\left(i\frac{x^2}{4t}\frac{m}{\hbar}\right) w\left(\frac{-x}{\sqrt{-2it\hbar/m}}\right)\sqrt{n} \qquad (3)$$

We can therefore calculate the current density

$$j(x,t) = \frac{\hbar}{m}\Im\left(\psi^* \frac{\partial \psi}{\partial x}\right) = \frac{\hbar n}{2m}\Im\left[\frac{1}{\sqrt{2it\pi\hbar/m}} w\left(-\frac{x}{\sqrt{2it\hbar/m}}\right)\right] \qquad (4)$$

where $\Im$ stands for the imaginary part.

It is important to point out that since the initial wavefunction was a real function (accept maybe for an arbitrary phase) then by its definition

$$j(x,-t) = -j(x,t) \text{ and evidently } j(x, t=0) = 0. \qquad (5)$$

In the short time approximation, where the measurement is taken far from the singularity, i.e., $t \ll 2mx^2/\hbar$

$$j(x,t) \to \frac{\hbar}{2m}\frac{n}{\pi x} \qquad (6)$$

i.e., at short time, the current is independent of time! This result is particularly strange since initially the current must vanish for $x > 0$. More accurately, due to (5)

$$j\left(x, |t| \ll \frac{2mx^2}{\hbar}\right) \to \begin{cases} \hbar n/2m\pi x & t > 0 \\ 0 & t = 0 \\ -\hbar n/2m\pi x & t < 0 \end{cases}$$

Therefore, the current jumps from 0 to $j = \hbar n / 2m\pi x$ instantaneously! Obviously, in the smooth\continuous scenario (see below) this instantaneous conduct disappears. The particles density, on the other hand, is continuous. For any $t$:

$$\rho = |\psi|^2 = \frac{1}{4}\left|w\left(\frac{-x}{\sqrt{-2it\hbar/m}}\right)\right|^2 n,$$

which vanishes at the limit $t \to +0$ like (see also Ref.8):

$$\rho(x,t) \sim \frac{\hbar}{2m}\frac{|t|}{\pi x^2}n \qquad (7)$$

Unlike the current $\rho(x,-t) = \rho(x,t)$.

Evidently, the continuity equation holds

$$\frac{\partial \rho}{\partial t} + \frac{\partial j}{\partial x} = 0,$$

but we see the peculiar behavior that the particles density vanishes while the current does not.

The reason for this phenomenon is a combination of the discontinuity and the Schorodinger dynamics.

Clearly, the Schrodinger equation is not a causal differential equation; however, even if we solve causal systems, such as the Klein-Gordon equation[8-9], the Schrödinger approximation is valid with high fidelity for $t \gg x/c$, where c is the speed of light. That is, we can still limit the discussion to the temporal range $x/c \ll t \ll 2mx^2/\hbar$.

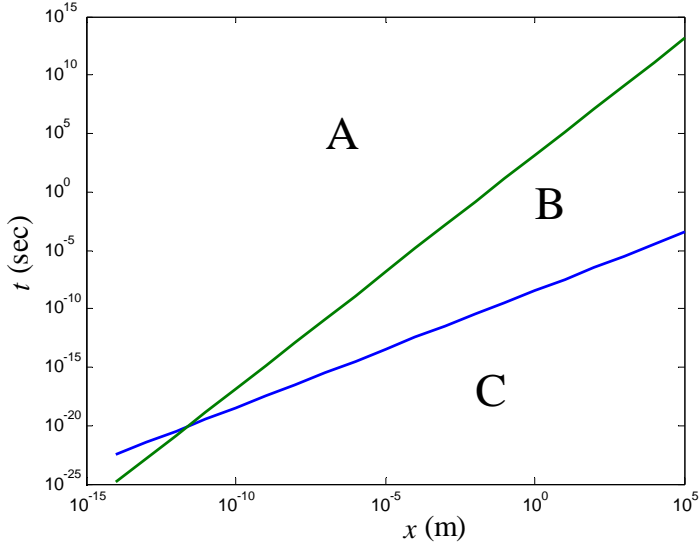

Fig.1: The validity regime of the short-time Schrödinger dynamics for electrons. A) ordinary Schrödinger dynamics, B) Short-time Schrödinger dynamics, and C) Relativistically non-causal dynamics.

This time period will be long enough when $x \gg \hbar/mc$. This is a very small length scale ($<10^{-12}m$ in the case of electrons, see Fig.1, and practically zero for atoms), much smaller than an atom and only two orders of magnitude larger than the classical size of an electron. Hence, we can say for sure that causality does not pose a limitation to witness the effect.

Nevertheless, the Schrödinger dynamics alone cannot explain this effect. For an initially smooth function the current always increases gradually with time. The discontinuity is essential for this instantaneous current.

Except for the momentum distribution, most of the effect can be explained by classical considerations.

**Semi-classical analog**: Consider a classical shutter beyond which ($x<0$) there is an infinite number of classical particles with a certain velocity distribution (Fig.2). To determine the distribution we need a quantum (wave-like) property of the particles.

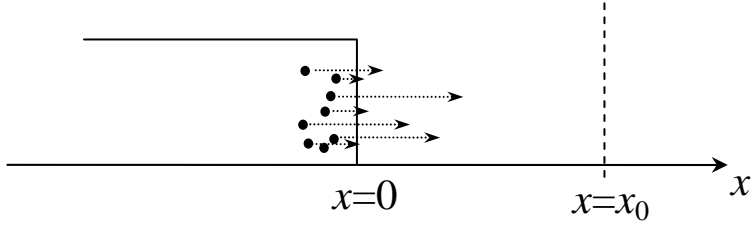

Fig.2: A semi-classical realization of the transient effect.

Since the Fourier transform of the initial wavefunction (the step function) goes like $\sim 1/k$, where $k$ is the particle's wavefunction, then the particles spectral distribution is proportional to $\sim 1/k^2$. Hence, the distribution of the number of particles, whose velocity exceeds $v_0$ [$N(v>v_0)$] is proportional to the reciprocal of $v_0$ (or momentum), i.e.,

$$N(v>v_0) \sim v_{max}/v_0. \tag{8}$$

where $v_{max}$ is assumed to be the maximum velocity in the ensemble. This is the only place in the argument, where the wave-like property of the particles is needed.

Then, at $t=0$ the shutter is removed and the particles propagate freely in space (and hence there is no change in the particles distribution). Now, if a particle is measured at a distance $x_0$ from the shutter (the singularity) at time $t$ then its velocity is exactly $x_0/t$.

Therefore, at time $t$ the number of particles, which reach a distance beyond $x_0$ goes like

$$N(t, x>x_0) = \int_{x_0}^{\infty} \rho(x)dx \sim v_{max}t/x_0. \tag{9}$$

Hence, the particles density, which is the spatial derivative of $N$ satisfies

$$\rho(x) \sim v_{max}t/x_0^2, \tag{10}$$

and finally the current density is indeed time independent

$$j(x) = \rho(x)v_0 \sim v_{max}/x_0. \tag{11}$$

These two last results agree with Eq.7 and Eq.5 respectively.

**The smooth/continuous case**:

Clearly, in a real experiment there is a certain width to the transition. If we take an initial wavefunction, which has a finite gradient $\xi$ like

$$\psi(x,t=0) = \frac{\sqrt{n}}{2}\mathbf{erfc}\left(\frac{x}{\xi}\right), \tag{12}$$

Clearly, this is only an example. Any smoothening of the step function will do.
We will find that this initial state has an exact analytical solution for any time $t > 0$:

$$\psi(x,t) = \frac{\sqrt{n}}{2}\mathbf{erfc}\left(\frac{x}{\sqrt{i2t\hbar/m + \xi^2}}\right) \tag{13}$$

and therefore, the probability and current densities are

$$\rho(x,t) = |\psi(x,t)|^2 = \frac{n}{4}\left|\mathbf{erfc}\left(\frac{x}{\sqrt{i2t\hbar/m + \xi^2}}\right)\right|^2 \tag{14}$$

and

$$j = \frac{\hbar}{m}\mathfrak{I}\left(\psi * \frac{\partial \psi}{\partial x}\right) = \\ -\frac{\hbar}{2m}\frac{n}{\sqrt{\pi}}\mathfrak{I}\left(\mathbf{erfc}\left(\frac{x}{\sqrt{-i2t\hbar/m + \xi^2}}\right)\exp\left(-\frac{x^2}{i2t\hbar/m + \xi^2}\right)\frac{1}{\sqrt{i2t\hbar/m + \xi^2}}\right) \tag{15}$$

respectively. In Fig.3 we plot the two. Despite the initial finite gradient $\xi$ the finite current temporal regime is clearly seen.

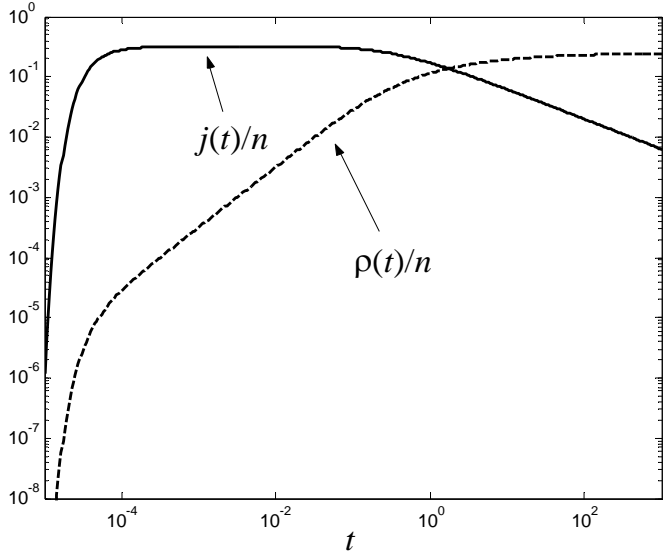

Fig. 3: The probability density and the current density for an initially finite width gradient. If $x$ is the arbitrary measurement place then $\xi/x = 1\times 10^{-4}$, the time is measured in units of $2mx^2/\hbar$, $\rho(t)/n$ is dimensionless and $j(t)/n$ is measured in units of $\hbar/2m$.

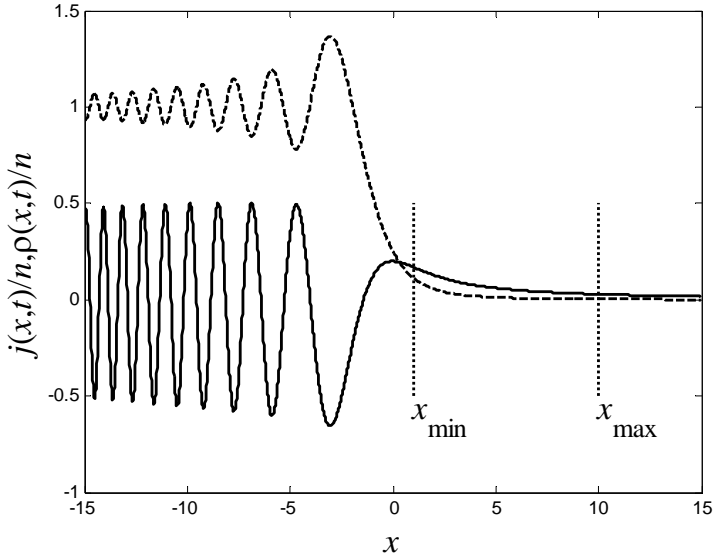

Fig.4: The exact probability density (dashed line, Eq.14) and the current density (solid line, Eq.15) are plotted vs. spatial coordinate $x$. For any arbitrary time scale $t$, $x$ is measured in units of $\sqrt{\hbar t/2m}$, the transient length scale $\xi = 0.1\sqrt{\hbar t/2m}$, $\rho(t)/n$ is dimensionless and $j(t)/n$ is measured in units of $\hbar/2m$. The time-independent current regime is marked between $x_{\min} = \sqrt{\hbar t/2m}$ and $x_{\max} = \hbar t/2m\xi$.

At short time ($t \ll 2mx^2/\hbar$) the wavefunction looks like:

$$\psi(x,t) \sim \frac{\sqrt{n}}{2\sqrt{\pi}} \frac{\sqrt{i2t\hbar/m + \xi^2}}{x} \exp\left(-\frac{x^2}{i2t\hbar/m + \xi^2}\right), \tag{16}$$

which yield the following probability and current densities:

$$\rho(x,t) \sim \frac{n}{4\pi} \frac{\sqrt{(2t\hbar/m)^2 + \xi^4}}{x^2} \exp\left(-\frac{2\xi^2 x^2}{(2t\hbar/m)^2 + \xi^4}\right) \tag{17}$$

and

$$j \sim \frac{\hbar}{2m} \frac{n}{\pi x} \frac{2t\hbar/m}{\sqrt{(2t\hbar/m)^2 + \xi^4}} \exp\left(-\frac{2\xi^2 x^2}{(2t\hbar/m)^2 + \xi^4}\right) \tag{18}$$

respectively.

Thus, they have a simple relation between them

$$j \sim \frac{8xt\hbar/m}{(2t\hbar/m)^2 + \xi^4} \rho(x,t). \tag{19}$$

As can be seen from (16)-(19), what we called "short time" can be divided into shorter periods. Actually, the constant current appears only in the temporal regime

$$x\xi \frac{2m}{\hbar} \ll t \ll x^2 \frac{2m}{\hbar}.$$

Before the time scale $t_0 = x\xi(2m/\hbar)$ the current increases gradually.

Initially ($t \ll 2m\xi^2/\hbar$) the current increases linearly with time as one would expect in an ordinary (i.e., continuous) Schrödinger process

$$j \sim \left(\frac{\hbar}{m}\right)^2 \frac{n}{\pi x} \frac{t}{\xi^2} \exp\left[-2(x/\xi)^2\right] \tag{20}$$

In this initial period the particles density remains in its initial condition, i.e.,

$$\rho(x,t) \sim \frac{n}{4\pi} \frac{\xi^2}{x^2} \exp\left[-2(x/\xi)^2\right], \tag{21}$$

and the ratio between them (which usually has the meaning of velocity) is a product of the spatial and temporal measurement locations

$$\frac{j}{\rho(x,t)} \sim \left(\frac{\hbar}{m}\right)^2 \frac{4xt}{\xi^4}. \tag{22}$$

However this temporal regime is usually too short to measure (we will discuss that in the next section).

During the next interval (i.e., beyond the first time scale) $2m\xi^2/\hbar \ll t \ll 2m\xi x/\hbar$ the particle density as well as the current increases exponentially

$$\rho(x,t) \sim \frac{\hbar}{2m} \frac{n}{\pi} \frac{t}{x^2} \exp\left(-\frac{\xi^2 x^2}{2(t\hbar/m)^2}\right) \tag{23}$$

and

$$j \sim \frac{\hbar}{2m} \frac{n}{\pi x} \exp\left(-\frac{\xi^2 x^2}{2(t\hbar/m)^2}\right) \tag{24}$$

respectively.

Since at this period, most of the particles at the measurement location originally arrived from the singularity then there velocity is approximately $x/t$ and then

$$j \sim \frac{x}{t}\rho(x,t). \tag{25}$$

It takes approximately a time period of $t_0 = x\xi(2m/\hbar)$ until the current saturates (Eqs.6 and 7), after which it remains unchanged for a very long period (until $t \approx x^2(2m/\hbar)$)

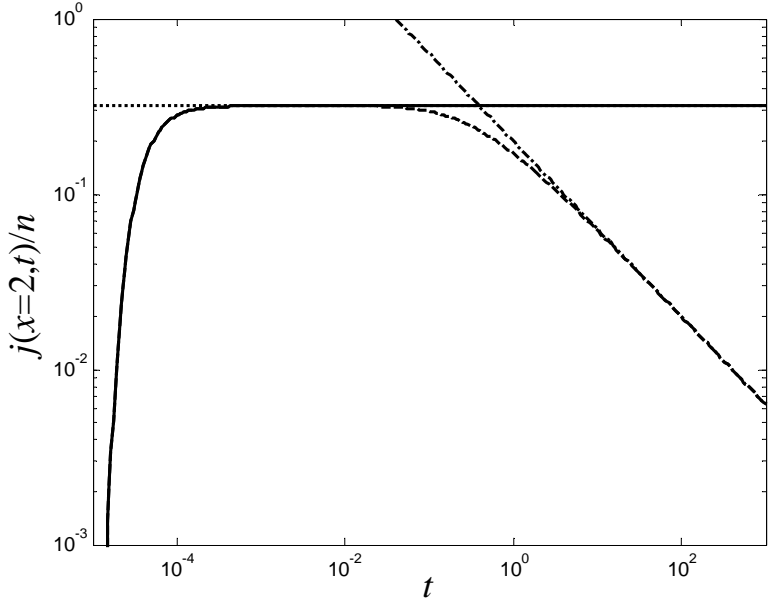

Fig.5: The different temporal regimes of the current. The dashed line is the exact solution (Eq.15), the solid line is the short time approximation (Eq. 24), the dash-dot is the long time approximation (Eq.26), and the dotted line is the time-independent approximation (Eq.5). The axes units and the parameters are the same as for Fig.3

After this stationary current period ($t \gg 2mx^2/\hbar$) the current decays very slowly

$$j \sim \sqrt{\frac{\hbar}{m\pi t}} \frac{n}{4} \qquad (26)$$

while the particles density converges to

$$\rho(x,t) \to \frac{n}{4}. \qquad (27)$$

Now we have to full picture. In the singular/discontinuous case the constant current (Eq.6) emerges instantaneously for $t > 0$ and last till $t \sim 2mx^2/\hbar$. However, when the initial wavefunction is a smooth gradient with $\xi$ as the transient scale, then the current increases gradually from 0 to its time-independent value (7) $(\hbar/2\pi m)(n/x)$ within the period $t_0 = x\xi(2m/\hbar)$, beyond which the current remains constant until $t \sim 2mx^2/\hbar$. Therefore, the temporal regime, where this constant current appear is

$$\frac{m}{\hbar} x\xi \ll t \ll \frac{m}{\hbar} x^2 \qquad (28)$$

**Possible Experimental Realization**

The initial wavefunction (12) can be constructed as an eigenstate of the potential

$$V(x) = \frac{2}{\sqrt{\pi}} \frac{\hbar^2}{m\xi^2} \frac{x}{\xi} \frac{\exp[-(x/\xi)^2]}{\text{erfc}(x/\xi)}$$

However, any potential, which has a socket in front of the boundary, whose width is $\sim \xi$ and depth is $\sim \xi^{-2}$, will do. As a particular case, laser atom trapping[10-15] allow us to sculpture the shape of the trapping potential.

If the particles are cooled Rubidium atoms[16-17], which are trapped in a laser beam then a multi transversal modes Gaussian beam, whose transversal decay length is $\xi$ can simulate such a potential for $t < 0$. Then, at $t = 0$ the beam is turned off and the particles current is measured a distance $x$ from the trap. Then, if $\xi$ is measured in nanometers, and $x$ is measured in meters then the period in which we have a chance of seeing this constant current is

$$\left(\frac{x}{1m}\right)\frac{\xi}{1nm} \ll \frac{t}{1\text{sec}} \ll 10^9 \left(\frac{x}{1m}\right)^2. \tag{29}$$

For example, if the transition (i.e., the optical trap boundaries) takes place at $\xi \cong 100nm$ and the current is measured a distance of $\cong 1cm$ from the trap then

$$1\text{sec} \ll t \ll 24\text{hours} \tag{30}$$

Clearly, for most cases, these results suggest a constant current throughout the entire experiment.

**Summary**

We have analyzed the Schrödinger dynamics of an initially high gradient wavefunction. It is shown that in this case a current emerges almost instantenuously even when the particles density is arbitrarily small, then, the current remains unchanged during all the "short-time" period despite the high transient. We suggest an experimental scenario with initially trapped cooled atoms to witness the time-independent current effect.